# Surface roughness influence on parametric amplification of nanoresonators in presence of thermomechanical and environmental noise

G. Palasantzas [a]

Zernike Institute of Advanced Materials, University of Groningen, Nijneborgh 4, 9747 AG Groningen, The Netherlands

**Abstract**

We investigate the surface roughness influence on the gain from parametric amplification in nanoresonators in presence of thermomechanical and momentum exchange noise. The roughness is characterized by the rms amplitude *w*, the correlation length $\xi$, and the roughness exponent $0<H<1$. It is found that the gain strongly increases with increasing roughening (decreasing *H* and/or increasing ratio $w/\xi$) due to increment of capacitive coupling, which plays dominant role when the intrinsic quality factor $Q_{in}$ is comparable or lower than the quality factor $Q_{gas}$ due to gas collisions. However, for $Q_{in}>>Q_{gas}$, the influence of surface roughness on the gain strongly diminishes.

*Pacs numbers:* 85.85.+j, 73.50.Td, 68.55.-a, 74.62.Fj

[a]Corresponding author: G.Palasantzas@rug.nl

The detection of small forces and the mass of molecules adsorbed on surfaces are of interest in a wide area of research fields, so as scanning probe microscopy, gravity wave detection, and mass sensor technologies [1-3]. Typically, the detection includes the conversion of the mechanical motion to an electrical signal via a transducer and then amplifying the electrical signal. A mechanical parametric amplifier greatly improves the mechanical response of micro/nanocantilevers responding to small harmonic forces [4]. In this case, the possibility of squeezed thermomechanical noise has also been demonstrated [4]. Recently it was also shown that mass sensing at long averaging times, where the noise close to the carrier is sampled, the Allan variance (which gives the limit to mass sensitivity) becomes worse with increasing parametric amplification [5]. Indeed, it was shown that the parametric amplification did not give improved performance over that achieved in the linear regime [5].

Nonetheless, it remains unexplored how parametric amplification will be influenced by environmental noise due to impingement of gas molecules [6], besides thermomecanical noise, if the resonator surfaces are rough. This source of noise is influenced by the surface morphology of the oscillating cantilever [7], which is a possibility to be further explored for parametric sensing. Notably, the influence of surfaces on nanoelectromechanical systems (NEMS) has been shown in variety of studies. Indeed, NEMS of SiC/Si have been shown to be operational in the UHF/microwave regime when having low surface roughness, while devices with rougher surfaces could not be operated higher than the VHF regime [8]. Studies of Si nanowires have shown the quality factor to decrease by increasing surface area to volume ratio [9]. Recently random surface roughness was also shown to affect the quality factor, dynamic range, adsorption-desorption noise, and limit to mass sensitivity of nanoresonators [7, 10]. Therefore, these considerations motivate the present work to explore how surface dependent fluctuation processes can influence parametric amplification.

We base our calculation on the simple harmonic oscillator model for the cantilever vibration $u(t)$ for its free end, having an effective oscillating mass $M_{eff}$ and spring constant $k_{eff}$ that results in a resonance frequency $\omega_o = (k_{eff}/M_{eff})^{1/2}$. For parametric sensing a modulation of the spring constant $k(t)$ at a frequency $2\omega_o$ is added, which is controlled using, e.g., a capacitive coupling between the cantilever on top of a fixed electrode [4]. The advantage of parametric sensing is that as $k(t)$ is increased in amplitude, the response of the resonator to a weak external driving force $F(t)$ (representing the signal to be detected) is significantly amplified for drive frequencies near to $\omega_o$ [4, 5].

The detected force is assumed to have the form $F(t) = F_o \cos(\omega_o t + \varphi)$ with $\varphi$ the phase angle between this external modulation and the independent actuating drive, and the pump voltage the form $\tilde{V}(t) = V_o + V \sin(2\omega_o t)$ (representing a particular degenerate case for this choice of frequencies). The latter yields a time varying spring constant $k(t) = k_c \sin(2\omega_o t)$ with $k_c = V_o V (\partial^2 C / \partial x^2)$ and $C(x)$ the capacitance between cantilever and ground electrode [4]. The resonator's equation of motion has the form [4, 5]

$$M_{eff} \ddot{u} + [k_{eff} + k(t)]u = -(M_{eff}\omega_o/Q_{in})\dot{u} - (M_{eff}\omega_o/Q_{gas,r})\dot{u} + F(t). \qquad (1)$$

$-(M_{eff}\omega_o/Q_{in})\dot{u}$ is the force due to the intrinsic damping associated with thermomechanical noise due to coupling between the resonator and its dissipative reservoir with intrinsic quality factor $Q_{in}$. The resonator can also undergo gas damping due to impingement and momentum exchange of gas molecules on its surface [6, 7] yielding a drag force $-(M_{eff}\omega_o/Q_{gas,r})\dot{u}$ with quality factor $Q_{gas,r} = M_{eff}\omega_o \sqrt{K_B T/m}(PA_{rou})^{-1}$. $P$

is the gas pressure, $m$ is the molecule mass, and $A_{rou}$ is the rough surface area of the beam [6, 7]. Here we assumed that the resonator operates within the molecular regime or molecule mean free path $L_{mph}$ larger than the beam width $w_b$ (<< beam length $L$) [11]. The gain is defined as the maximum displacement amplitude for $V>0$ divided by that for $V=0$ and it given as in (below the threshold for self-sustained oscillations) [4] by

$$Gain = \{\cos^2 \varphi (1+V/V_t)^{-2} + \sin^2 \varphi (1-V/V_t)^{-2}\}^{1/2}, \qquad (2)$$

where $V_t = 2k_{eff}/QV_o(\partial^2 C/\partial x^2)$. Q is the total quality factor: $1/Q = 1/Q_{in} + 1/Q_{gas,r}$.

In order to further compute the gain we have to compute Q and thus $Q_{gas,r}$ due to gas dissipation. If we assume for the roughness profile of the resonator surface a single valued random function $h(r)$ of the in-plane position $r=(x,y)$ and a Gaussian height distribution [12], the rough area is given by $A_{rou}/A_{flat} = R_{rou} = \int_0^{+\infty} du \left(\sqrt{1+\rho^2 u}\right) e^{-u}$ [13] with $\rho = <\sqrt{<(\nabla h)^2>}> = (\int_{0 \leq q \leq Q_c} q^2 <|h(q)|^2> d^2q)^{1/2}$ the average local surface slope [14], and $A_{flat} = w_b L$. $<|h(q)|^2>$ is the roughness spectrum. $Q_c = \pi/a_o$ with $a_o$ a lower lateral cut-off. If we substitute in Eq. (2), and take into account the fact that the capacitance $C(x)$ is proportional to surface area and thus to $R_{rou}$, $C \sim R_{rou}$ (yielding $\partial^2 C/\partial x^2 \approx R_{rou} \partial^2 C_{flat}/\partial x^2$) one obtains for the gain the final form

$$Gain \approx \{\cos^2 \varphi [1 + (V/V_{t,in} R_{rou}^{-1}(1+Q_{in} R_{rou}/Q_{gas,f}))]^{-2} \\ + \sin^2 \varphi [1 - (V/V_{t,in} R_{rou}^{-1}(1+Q_{in} R_{rou}/Q_{gas,f}))]^{-2}\}^{1/2} \qquad (3)$$

where $V_{t,in} = Q_{in}^{-1}[2k_{eff}/V_o(\partial^2 C_{flat}/\partial x^2)]$ and $Q_{gas,f} = M_{eff} \omega_o \sqrt{K_B T/m} (PA_{flat})^{-1}$.

The gain calculations in terms of Eq. (3) were performed for random self-affine rough surfaces, which it is observed in a wide spectrum of surface engineering processes [12]. In this case $<|h(q)|^2>$ scales as $<|h(q)|^2> \propto q^{-2-2H}$ if $q\xi>>1$, and $<|h(q)|^2> \propto const$ if $q\xi<<1$ [12, 15]. This is satisfied by the analytic model [15] $<|h(q)|^2> = (2\pi w^2 \xi^2)/(1+aq^2\xi^2)^{(1+H)}$ with $a = (1/2H)[1-(1+aQ_c^2\xi^2)^{-H}]$ if $0<H<1$, and $a = 1/2\ln(1+aQ_c^2\xi^2)$ if $H=0$. Small values of $H$ (~0) characterize jagged or irregular surfaces; while large values of $H$ (~1) surfaces with smooth hills-valleys [12, 15]. In addition, we obtain for the local slope the analytic expression $\rho = (w/\sqrt{2}\xi a)\{(1-H)^{-1}[(1+aQ_c^2\xi^2)^{1-H}-1]-2a\}^{1/2}$ [14], which further facilitates calculations of the gain from Eq. (3). For other roughness models see ref. [16].

The numerical calculations were performed for roughness amplitudes observed in real nanoresonator surfaces $w$~3-10 nm [8], and for the cut-off we used $a_o$=0.3 nm. Figures 1 and 2 shows calculations of the gain for two different phase angles $\varphi$ from close to the deamplifying regime $\varphi$~0 up to full amplification that occurs for $\varphi=90^0$. Even for small $\varphi$, some amplification can occur for pump voltages $V$ close to the threshold value $V_t$ due to contribution from the amplifying term (proportional to $\sin^2\varphi$) in Eq. (3). Moreover, as Fig. 3 indicates in comparison with Fig. 1, with decreasing gas quality factor $Q_{gas,f}$ (so that $Q_{gas,f}<Q_{in}$) the amplification occurs at larger pumping voltages $V$ since in this case $V_t$ increases significantly. Therefore, the gas environment plays significant role on parametric amplification.

Furthermore, Figs.1-3 shows calculations of the gain for various roughness exponents $H$. If we compare Figs.1 and 2, then it becomes evident that the influence of surface roughening becomes more distinct for pump voltages $V$ close to $V_t$. Indeed, decreasing $H$ (equivalently increasing roughness at short length scales $<\xi$) leads to decrement of the critical voltage $V_t$ where maximum amplification occurs. Comparing Figs.1 and 3, one can conclude that the gain increases with increasing roughening (i.e., decreasing $H$ and/or increasing ratio $w/\xi$) due to increment of the capacitive coupling associated with increasing surface area, which however plays dominant role when the intrinsic quality factor is comparable or lower than that due to gas collisions. In the opposite case, $Q_{in} >> Q_{gas,f}$, the influence of surface roughness on the gain strongly diminishes because the increment of gas dissipation and the increment of capacitive coupling counterbalance each other. Indeed, from Eq. (3) we obtain for $Q_{in} >> Q_{gas,f}$

$$Gain \approx \{\cos^2 \varphi [1+(V/V_{t,in})(Q_{gas,f}/Q_{in})]^{-2} \\ + \sin^2 \varphi [1-(V/V_{t,in})(Q_{gas,f}/Q_{in})]^{-2}\}^{1/2}, \qquad (4)$$

which indicates that the gain does not depend on surface roughness.

Finally, if we compare Figs. 1 and 2, it is evident that for large phase angles $\varphi$ (close to the maximum amplifying regime $\sim 90^o$) increasing roughness leads to higher amplification, while for small angles $\varphi$ (close to the deamplifying value $\varphi=0$) the influence of surface morphology is not significant for $V<<V_t$. For clarity, Fig. 4 shows the direct plot of the *Gain* vs. phase angle $\varphi$ for various exponents $H$ and a sufficiently large pumping voltage $V$ (comparable to $V_t$). As it is shown in Fig. 4 increasing roughness (i.e., decreasing $H$ in these schematics) leads to higher amplification for phase angles $\varphi \sim 90^o$, while for phase values close to $\varphi \sim 0^o$ it leads to higher de-amplification. Similar is the effect of increasing roughness amplitude $w$ and/or decreasing correlation length $\xi$ (see inset in Fig.4) since in both cases surface roughening occurs.

In conclusion, we investigated the simultaneous influence of thermomechanical and momentum exchange noise on the gain of non-linear parametric amplification. It is found that the amplification gain strongly increases with increasing roughening due to increment of capacitive coupling, which plays dominant role when the intrinsic quality factor is comparable or lower than the quality factor due to gas collisions. This result will hold qualitatively also for non self-affine roughness models as long as variations of the characteristic roughness parameters leads to rougher surfaces. In the opposite case the influence of surface roughness is negligible. Notably, these considerations should be taken into account in real parametric amplifiers with nanoscale rough surfaces.

**Acknowledgements:** I would like to thank A. Cleland for useful communication.

**Figure Captions**

**Figure 1** Gain vs. pumping voltage *V* for phase angles $\varphi=90^o$, *H* as indicated, $\xi=60$ *nm*, *w=3 nm*, $Q_{in}/Q_{gas,f}=1$, and $Q_{in}=10^4$.

**Figure 2** Gain vs. pumping voltage *V* for phase angles $\varphi=10^o$, *H* as indicated, $\xi=60$ *nm*, *w=3 nm*, $Q_{in}/Q_{gas,f}=1$, and $Q_{in}=10^4$.

**Figure 3** Gain vs. pumping voltage *V* for phase angles $\varphi=90^o$, *H* as indicated, $\xi=60$ *nm*, *w=3 nm*, $Q_{in}/Q_{gas,f}=10$, and $Q_{in}=10^4$.

**Figure 4** Gain vs. phase angle $\varphi$, pumping voltage *V=1* Volt, *H* as indicated, $\xi=60$ *nm*, *w=3 nm*, $Q_{in}/Q_{gas,f}=1$, and $Q_{in}=10^4$. The inset shows similar calculations for *H=0.6* and different amplitudes *w* as indicated.

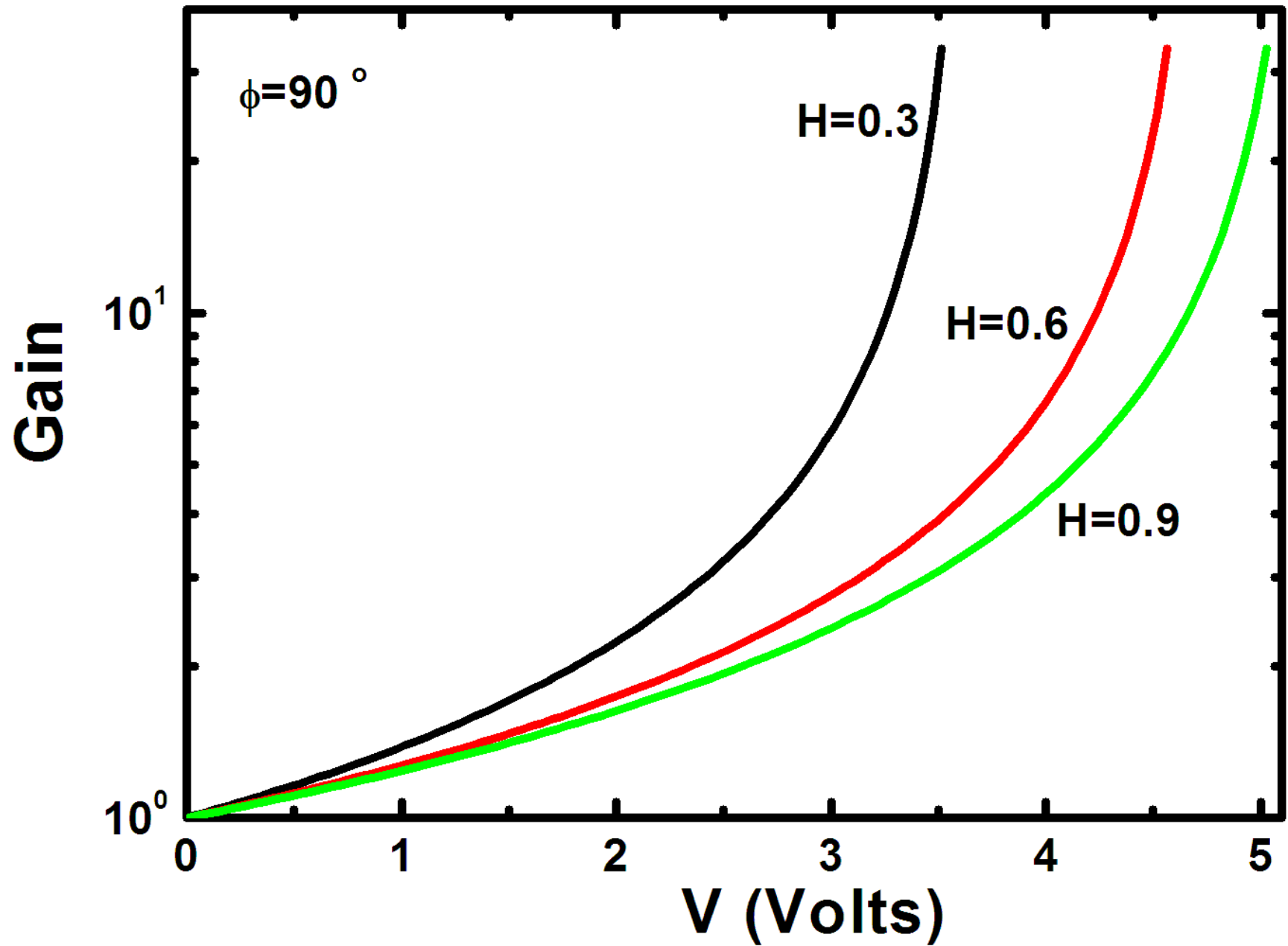

FIGURE 1

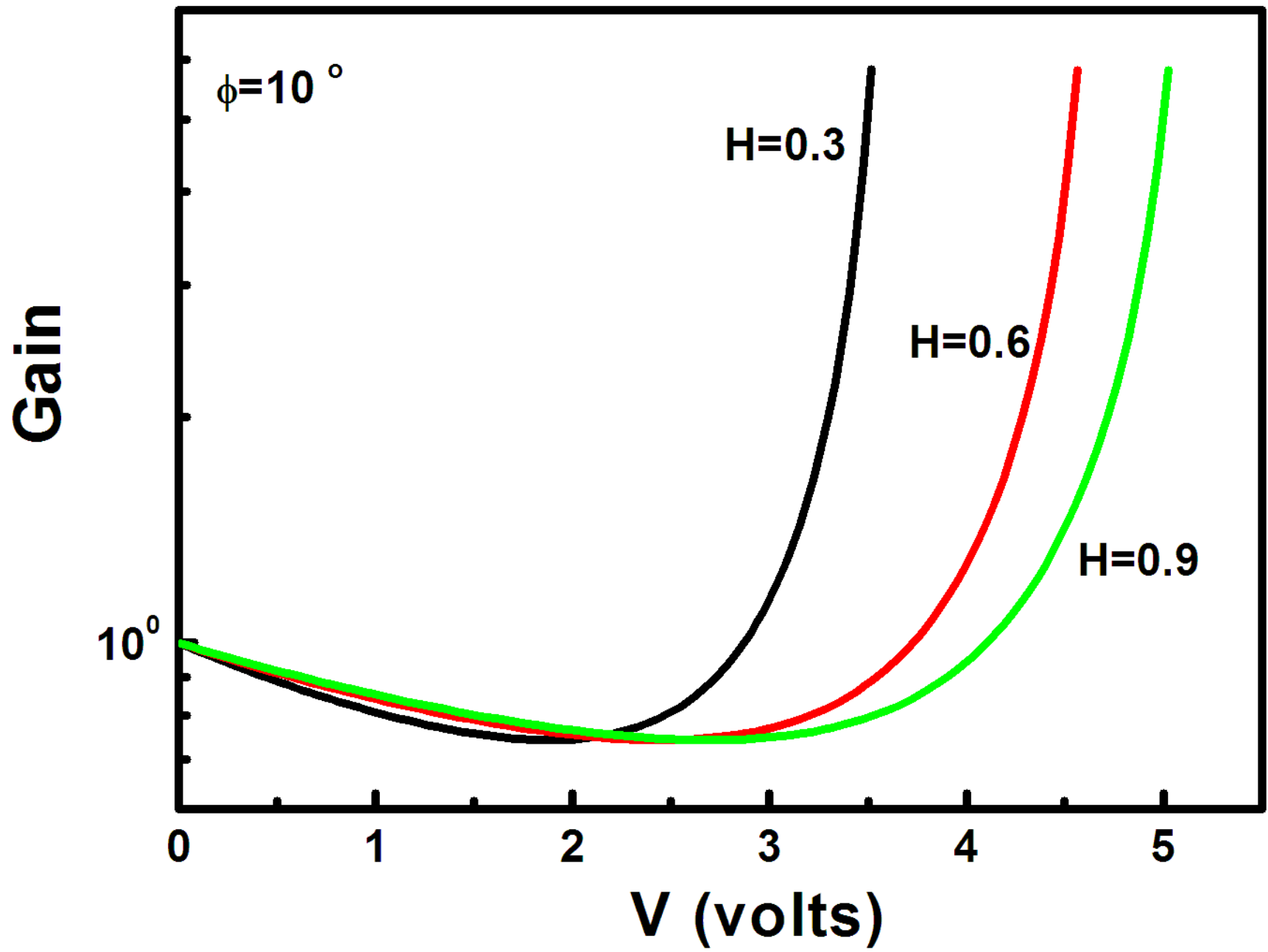

FIGURE 2

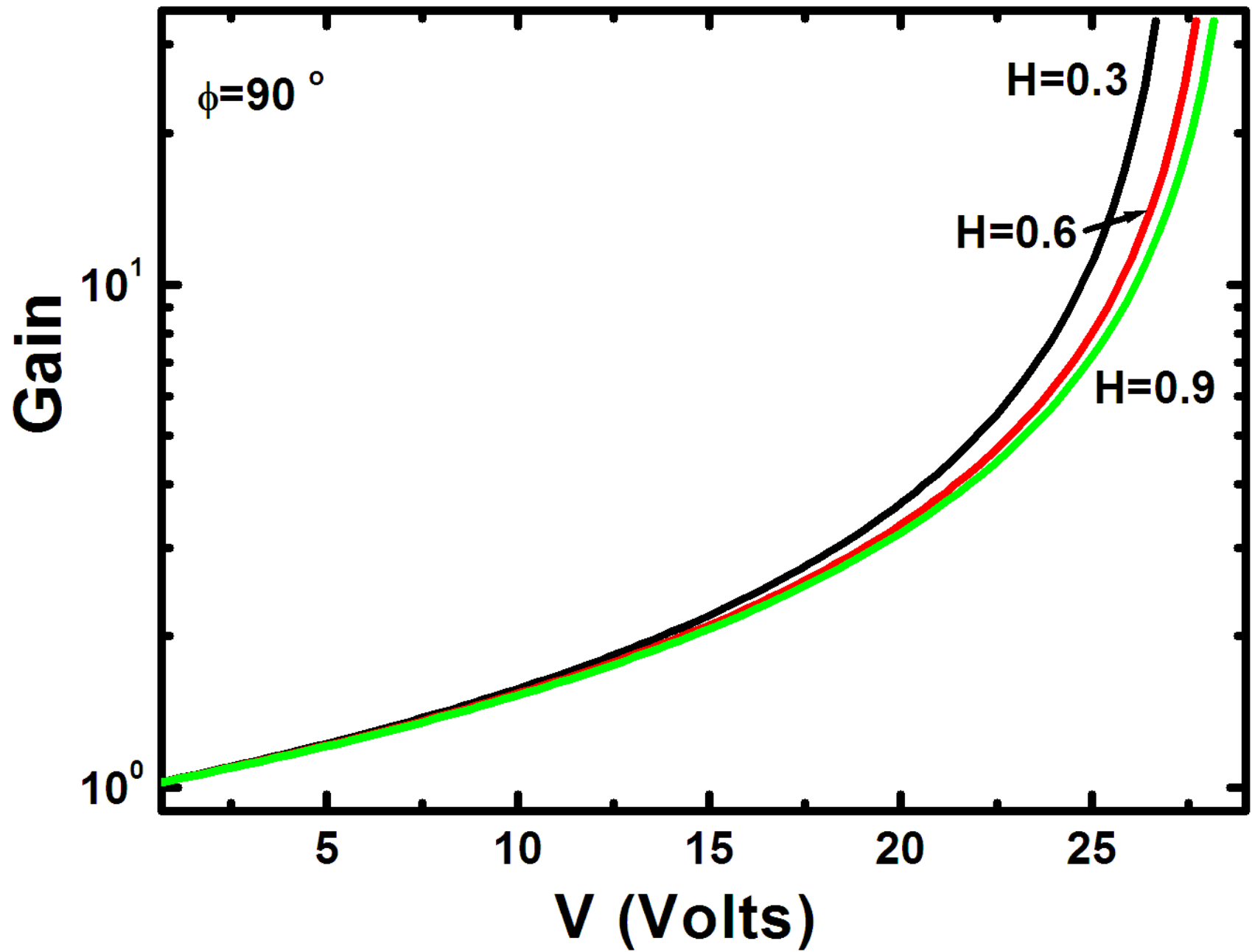

FIGURE 3

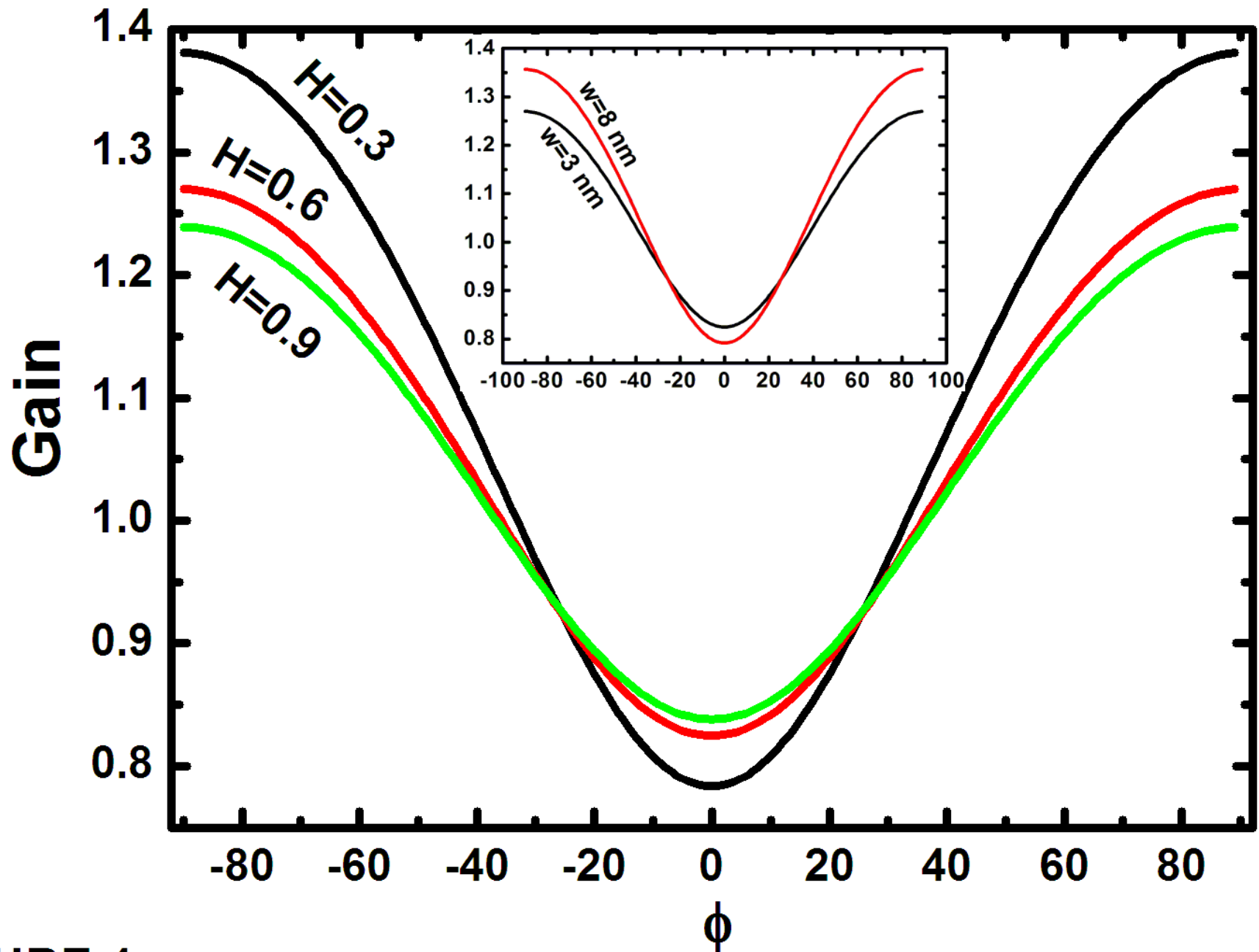

FIGURE 4